%% file: 0main.tex
\documentclass[letterpaper]{article}
\usepackage{tcolorbox}
\usepackage{url}
\usepackage{adjustbox}
\usepackage{multirow}
\usepackage{booktabs}
\usepackage{tabularx}
\usepackage{longtable}
\usepackage{rotating}
\usepackage{xspace}
\let\comment\relax

\newcommand{\plus}{\raisebox{.4\height}{\scalebox{.6}{+}}}

\newcommand{\comment}[1]{}

\newcommand{\remark}[1]{}

\newcommand{\ignore}[1]{}
\newcommand*{\codefont}{\fontseries{}\selectfont\texttt}

\usepackage{algorithm}
\usepackage{algpseudocode}
\usepackage{listings}
\usepackage{flushend}
\usepackage{authblk}
\usepackage{caption,subcaption}
\title{Replacements and Replaceables: Making the Case for Code Variants}

\author[1]{Venkatesh Vinayakarao}
\author[1]{Rahul Purandare}
\author[1]{Sumit Keswani}
\author[1]{Devika Sondhi}
\author[2]{Anita Sarma}
\affil[1]{IIIT-Delhi, India}
\affil[2]{Oregon State University, Corvallis, OR, USA.}

\begin{document}
\maketitle
\input{1_abstract}


\input{2_introduction}

\input{3_definitions}
\input{4_background}

\input{5_programvariants}

\input{6_experiments}
\input{61_applications}
\input{7_relatedwork}
\input{8_threats}
\input{9_conclusion}
\section{Acknowledgments}

\input{9_references}
\end{document}

%% file: 1_abstract.tex
\begin{abstract}
There are often multiple ways to implement the same requirement in source code. Different implementation choices can result in code snippets that are similar, and have been defined in multiple ways: \textit{code clones, examples, simions} and \textit{variants}. Currently, there is a lack of a consistent and unambiguous definition of such types of code snippets. Here we present a characterization study of code variants -- a specific type of code snippets that differ from each other by at least one desired property, within a given code context. We distinguish code variants  from other types of redundancies in source code, and demonstrate the significant role that they play: about 25\% to 43\% of developer discussions (in a set of nine open source projects) were about variants. We characterize different types of variants based on their code context and desired properties. As a demonstration of the possible use of our characterization of code variants, we show how search results can be ranked based on a desired property (e.g., speed of execution). 
\end{abstract}

%% file: 2_introduction.tex
\section{Introduction}
\label{IntroductionSection}
\comment{What are code variants?}\comment{Are they same as semantic clones?} 
``\textit{There's more than one way to skin a cat}", a popular English idiom, also holds true for software development. Developers often have to reason among the benefits and drawbacks of different implementation choices before making a selection. These choices can include differences in the speed or complexity of computation, the style of coding, the library used or licensing requirements. Currently, code snippets reflecting different choices are known by different names: variants, clones, simions, idioms, examples; and are often used interchangeably. In this work, we define and characterize code variants, and differentiate them from clones, simions, and examples.


\ignore{
Research literature points to various types of semantic similarities~\cite{Allamanis:2014:MIS:2635868.2635901,Gabel:2010:SUS:1882291.1882315}. For instance, semantic clones are similar to code variants due to the fact that both of them have behavioral similarities. However, developers seek code variants because of the need for some desired properties whereas they look to refactor and clean-up clones. Thus, clones and variants are two fundamentally different entities. Moreover, we observe that the definition of semantic clones is inconsistent across the research literature. }


\comment{Why code variants?} 
Code variants represent an alternative implementation of a code snippet, where each alternative provides the same functionality, but has different properties that make some better suited to the overall project requirements. 

Developers routinely need to analyze existing code, find better reuse alternatives, and look to develop high-quality code that meets some desired properties. For example, developers in Apache Commons Math discuss a variant in the FastMath library, which is faster, but is so because it is less accurate (Table~\ref{tbl:discussion} presents such an exchange in the issue tracker). The discussion involves issues about specific implementation choices, the library, the underlying algorithm, and license restrictions. Such discussions on selection of a specific variant are common. 

\begin{table}[]
\small
\centering
\caption{Code variants are discussed in the bug descriptions of popular open source projects.}
\label{tbl:discussion}
\vspace*{-5px}
\begin{tabular}{p{2.5cm}|p{5.3cm}}
\toprule
\textbf{Project}             & \textbf{Discussion}                                                                          \\
\midrule
Apache Math \scriptsize{[Bug\# MATH-901]} & \textit{One of the reasons this \textbf{variant} is faster is because it is less accurate, which may not be acceptable for commons-math.}  \\ \hline
Apache Math \scriptsize{[Bug\# MATH-1293]} & \textit{While the jury is still out, I made another \textbf{variant} of the patch ...}\\
\bottomrule
\end{tabular}
\vspace*{-10px}
\end{table}

\ignore{
 \hline
Thrift [Bug\# THRIFT-3217] & \textit{Provide a little endian \textbf{variant} of the binary protocol in C++                      } \\ }


\comment{Why variant mining tools?} Finding variants, however, is nontrivial. Developers search for variants over the web, in online discussion forums, and issue trackers. For example, when investigating Stack Overflow\footnote{http://stackoverflow.com/tour} (SO), a popular programming Q\&A site containing 13M questions and 20M answers, we found in a random sampling of 300 posts, 100 variants that discussed 34 topics and spanned 13 programming languages. Similarly, we found that 25\% to 43\% of developer discussions are about variants in our dataset. Our dataset comprised of fifteen open source projects in five different programming languages, with varying sizes and domains. However, searching for variants is  a fragmented, ad hoc process that can be time-consuming. 

       
\comment{Why is it difficult to develop them?}Mining for variants is an intrinsically difficult problem because of the following reasons. First, there is no consensus about what variants are. Research points to different types of semantic similarities~\cite{Allamanis:2014:MIS:2635868.2635901,Gabel:2010:SUS:1882291.1882315}, but there are inconsistencies in their definitions (see Section~\ref{Background}). Second, Juergens et al.~\cite{Juergens:2010:CSB:1955601.1955971} show that behaviorally similar code can vary significantly in structure, making it difficult to match source code patterns.
To the best of our knowledge, the concept of variants has not been discussed in the past and hence, it becomes important to understand what variants are and about its characteristics. Moreover, to build useful variant mining tools, we need to understand the code contexts and the desired properties that make a specific variant better than the rest. Our work attempts to draw some essential characteristics of variants which we believe could contribute towards automating variant mining which can help in the development process. 


\comment{Our contributions}To the best of our knowledge, ours is the first study in understanding variants and their characteristics. Our key contributions are as follows:
\begin{itemize}
\item We define and discuss how variants differ from semantic clones, simions and examples.
\item We study variants and characterize their code contexts and desired properties.
\item We analyze the source code and developer discussions in fifteen open source projects and show that variants exist and they matter to developers.
\item As an application of our variant characterization study, we propose a systematic approach to mine variants using data from discussion forums.
\end{itemize}

%% file: 3_definitions.tex
\section{Background and Terminology}
\label{Definitions}
\label{Background}
\begin{figure}[t]
\centering
\includegraphics*[trim=0cm 9cm 11.4cm 0cm, clip,width=8.6cm]{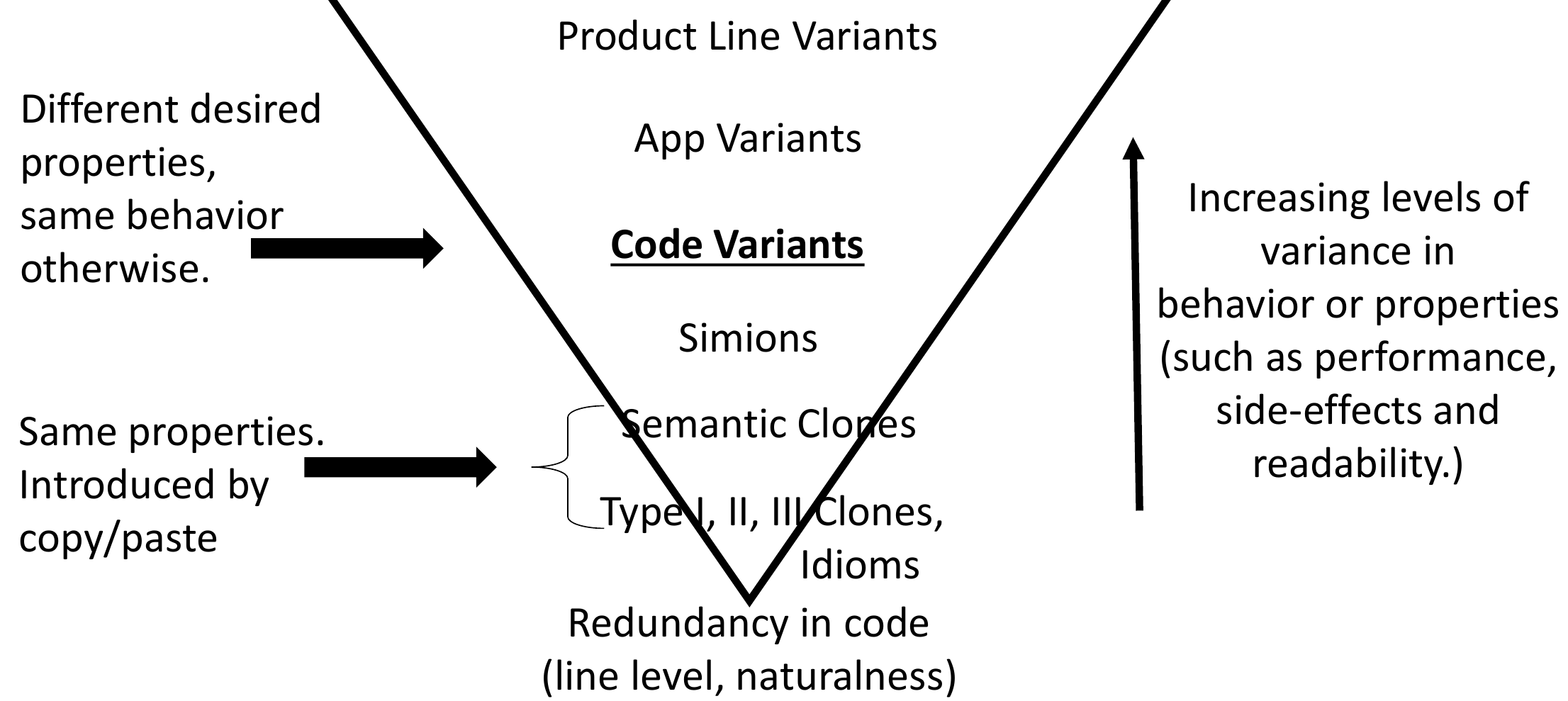}
\caption{Reuse in source code happens at different levels by size for different reasons. This work focuses on code variants.}
\label{fig:redundancy}
\vspace*{-10px}
\end{figure}

Reuse happens in source code at several levels. This phenomenon is captured in Figure~\ref{fig:redundancy}. Reuse starts at token-level, and occurs at line level too, as demonstrated by the naturalness of software~\cite{Hindle:2012:NS:2337223.2337322}. Idioms, clones, simions and code variants represent reuse over multiple lines of code. Also, we observe that reuse grows beyond programs into applications and product variants. This work focuses on reuse at the level of multiple lines of code.

Code variants are closely related to two major types of code snippets: 1) Semantically similar code snippets (clones, idioms and simions), and 2) Code examples. Each of these types has its similarities and dissimilarities with variants.

\comment{Background: What are program variants?}Before we compare variants with semantic clones, simions, idioms and program examples, we need a definition for them. We modify Muralidharan et al.'s~\cite{Muralidharan:2016:ACV:2872362.2872411} definition of variants as given below (and further discussed in Section~\ref{VariantsSec}):
\newtheorem{definition1}{Definition}
\begin{definition1}
A code variant represents an alternative implementation for a given code snippet under a specific context in which one of the two implementation choices must score better on at least one desired property over the other.
\end{definition1}

\ignore{Variants are discussed relative to two code snippets. Given two code snippets, we compare them within the given code context and check if one scores over the other on a desired property.} This emphasizes on \textit{code context} and \textit{desired properties} of code snippets. We adapt Kirke's~\cite{Krinke:2006:ECP:1167771.1167775} definition of method context to define code context as follows:

\begin{definition1}
Code context of a code snippet describes the fitment of the snippet inside the larger project. It captures the intent, dependencies of the surrounding code, input to the snippet, the output from the snippet and the states in which the system may get into.
\end{definition1}

In a variant pair, one variant is said to be \textit{preferred} over the other if it has at least one \textit{desired property} when compared with the other within the given context. As an example, Table~\ref{tbl:discussion} shows a trade-off between speed of execution and accuracy. In this context, accuracy is more important, and hence a variant which is more accurate is preferred. Therefore, we define properties as follows:

\begin{definition1}
\label{def:dprops}
Multiple implementation choices may satisfy the requirements in the given code context. Yet, each implementation choice (a code snippet) differs from one another in ways that could be either functional or para-functional or both. These qualities of code snippets that serve as differentiators are referred to as properties. 
\end{definition1}

Factors such as lack of knowledge may lead developers to use less optimal choices initially and later discover snippets with more desired properties. For example, a recursive version of factorial is seldom used in real-world projects especially if the return value cannot hold large values. A hard-coded version avoids recomputation. Table~\ref{tbl:variantsFact} shows these two versions of factorial implementation. If a developer goes for the less optimal choice inadvertently, it may eventually get changed during the review or maintenance process. Thus, the knowledge of variants at development time can help developers write better programs.

\begin{table}[t]
\centering
\caption{Two variants of factorial program are shown below. Note that (a) is a text book example for recursion. Since, the factorial value grows very large quickly, integer type can hold only up to 12!. Hence (b) is preferred for real-world projects since it avoids re-computation.}
\label{tbl:variantsFact}
\begin{tabular}{p{2cm}|p{4.2cm}}
\toprule
\textbf{(a) Recursive } & \textbf{(b) Real-World                                                                         } \\
\midrule
\begin{lstlisting}[belowskip=-0.8 \baselineskip, linewidth=3cm, aboveskip=-0.5\baselineskip]
public static int
 fact(int x) {
 if (x==1 | x==0)
   return 1;
 return fact(x-1)
             * x;
}
 \end{lstlisting}
 & \begin{lstlisting}[belowskip=-0.8 \baselineskip, aboveskip=-0.5\baselineskip]
public int factorial(int n) {
switch(n) { 
 case 0: return 1; 
  ... 
 case 12: return 
         479001600;
 default : throw new 
  IllegalArgumentException();}
} \end{lstlisting} \\
\bottomrule
\end{tabular}
\vspace*{-10px}
\end{table}


Next, we investigate the existing literature to differentiate variants from clones, simions, idioms and examples.

%% file: 4_background.tex
\subsection{Clones}
\comment{What are semantic clones?}Clones were originally defined as redundant snippets introduced due to a copy and paste activity~\cite{Juergens:2009:CCM:1555001.1555062}. They can be semantic or structural. Semantic clones appear in different varieties, such as wide-miss clones~\cite{Marcus:2001:IHC:872023.872542}, interleaved clones~\cite{Gabel:2008:SDS:1368088.1368132}, and high-level concept clones~\cite{Marcus:2001:IHC:872023.872542}.  Gabel et al.'s~\cite{Gabel:2008:SDS:1368088.1368132} definition of semantic clones is as follows: \textit{Two disjoint, possibly non-contiguous sequences of program syntax $S1$ and $S2$ are semantic code clones if and only if $S1$ and $S2$ are syntactic code clones or $\rho(S1)$ is isomorphic to $\rho(S2)$}. Here, $\rho$ is a Program Dependence Graph (PDG) based  transformation function. PDG captures control and data dependency in code snippets and abstracts away other syntactic details.  Elva and Leavens~\cite{Elva-Leavens12} define semantic clones as functionally identical code fragments. Ira Baxter defines clones as segments of code that are similar according to some (typically lexical) definition of similarity~\cite{DBLP:conf/dagstuhl/Koschke06}.



\comment{What are type-1,2,3 clones?}Syntactic clones~\cite{DBLP:conf/dagstuhl/Koschke06, 6240495} are of three types. Type-1 clones are exact copies. Type-2 clones are copies where only the identifier names and variable types are changed. They are otherwise structurally similar. There is no single accepted definition of Type-3 clones~\cite{DBLP:conf/dagstuhl/Koschke06, 6240495}. One definition of Type-3 clone is based on the Levenshtein distance between the pair of snippets which quantifies the minimum number of additions and deletions of tokens to transform one snippet to other. 


\subsection{Idioms and Simions}
Keivanloo et al.~\cite{6240495} and Juergens et al.~\cite{Juergens:2010:CSB:1955601.1955971} indicate that code similarities go beyond these definitions of clone types. Allamanis and Sutton~\cite{Allamanis:2014:MIS:2635868.2635901} define a code idiom as ``\textit{a syntactic fragment that recurs across software projects and serves a single semantic purpose}". They claim that programmers use the term idiomatic to refer to code that is used repetitively. Idioms have both syntactic and semantic similarity. 

Juergens et al.~\cite{Juergens:2010:CSB:1955601.1955971} call the snippets that are behaviorally similar as \textit{Simions}. Simions need not originate from copy and paste activity. Even though simions have independent origins, they have been described in the context of redundant code with the intent of identification and refactoring. 


\begin{table*}[ht]
\small
\centering
\caption{Fundamental differences exist between semantic clones, code examples and code variants.}
\label{tbl:CEVcomparision}
\vspace*{-10px}
\begin{tabular}{p{1.8cm}|p{3cm}|p{3cm}|p{3cm}}
\toprule
                        & \textbf{Semantic Clones} & \textbf{Code Examples} & \textbf{Code Variants} \\
                        \midrule
\textbf{Definition}              & 
Code snippets with no difference in properties of interest within the given code context. Therefore, one snippet can replace the other.  & Code snippets with an instructive property against an information need.             & Code variants represent alternative implementations suitable for a specific code context in which one variant must have some desired properties over the other.                \\
\textbf{Differences}             &  Clones are necessarily semantically similar and have no differences in desired properties. Hence, these snippets are redundant. Some amount of structural similarity is also assumed in cases where PDG based definitions are followed. &  Neither semantic nor syntactic similarity warranted.           Provides instructive value as in the usage of API or how to implement, and so on. &  Semantically similar but has different desired properties. \\   
\textbf{Example} & Two sorting implementations of same worst-case complexity where that is the only quality that matters to developers. & Any API usage tutorial. For example, in Java, the code snippets describing the usage of Arrays.sort feature. & Various sorting implementations with different time complexities are variants, if speed is the only desired property.
\\
\bottomrule    
\end{tabular}
\end{table*}

\subsection{Programming by Example}
\comment{Role of examples in development}Developers seldom read the entire documentation before they start. They learn from code snippets on the web or other projects~\cite{Panchenko:2011:DSS:1985429.1985438, Sahavechaphan:2006:XMS:1167473.1167508}. \textit{Code examples are small source code fragments whose purpose is to illustrate how a programming language construct, an API, or a specific function or method works}~\cite{Moreno:2015:IUT:2818754.2818860}. Examples play a significant role in comprehension, reuse, and bug-fixing~\cite{Sillito:2012:MGC:2473496.2473558}. As a result, several researchers have explored locating~\cite{Allamanis:2014:MIS:2635868.2635901,Brandt:2010:EPI:1753326.1753402}, selecting~\cite{McMillan:2010:RSC:1808920.1808925} and analyzing~\cite{Sillito:2012:MGC:2473496.2473558} examples.

%% file: 5_programvariants.tex
\section{Code Variants}
\label{VariantsSec}
\comment{Is the use of term ``variant" popular among developers? Is the use of term ``variant" popular in literature?} Use of the term ``variant" is quite popular in the development community. Table~\ref{tbl:variants} shows the number of occurrences of the term ``variant" in the same group of projects that Gabel et al. selected for semantic clones study. Not only developers but literature too provides several evidences of the use of the term ``variant"~\cite{Salamat:2009:OID:1519065.1519071,Muralidharan:2016:ACV:2872362.2872411}. Variants are used in at least three major contexts: 1) Code variants (focus of this paper), 2) Program or product variants (as in product lines and application variants)~\cite{Rubin:2013:FMC:2486788.2486971}, and 3) Configuration variants~\cite{Muralidharan:2016:ACV:2872362.2872411} (as in tuning a product or product configuration).

\paragraph*{Variants are neither clones, nor simions, nor idioms} 


Code variants are similar to clones in the sense that both are functionally similar set of code snippets. Variants differ from clones for the reasons of purpose and properties in the code context. Variants are always discussed with the intent of \textit{bringing in} code snippets with desired properties. Clones are discussed in the context of refactoring. Developers \textit{clean up} clones to promote reuse. Existence of a clone is considered as a bad smell. There is no difference between the desired properties present in the clone instances. 

Gabel's definition of clone does not capture these aspects of semantic clones. We suggest the following definition:

\begin{definition1}
\label{clonedefn}
Let $P$ be the set of $n$ desired properties $\{p_1,p_2,...,p_n\}$. Let $score_{p_i}(\phi)$ be a function computing the strength of code snippet $\phi$ over any property $p_i$. Code snippets $\nu_1$ and $\nu_2$ are \textbf{clones} if neither $\nu_1$ nor $\nu_2$ score over each other on any property of interest and thus $\nu_1$ and $\nu_2$ can replace each other in the given code context. In other words, $\forall_{i} ~score_{p_i}(\nu_1) = score_{p_i}(\nu_2)$.
\end{definition1}


This definition does not depend on structural similarity at all, and instead focuses on desired properties in a code context. This emphasis helps us differentiate variants from simions and idioms as well. Simions and idioms are semantically similar irrespective of the code context.

Unlike clones, a stronger variant scores overs a weaker variant at least on one desired property. Hence, we rephrase variants defined in Section~\ref{Definitions} as follows:

\begin{definition1}
\label{variantdefn}
Code snippets $\nu_1$ and $\nu_2$ are \textbf{variants} if there is at least one property of interest by which $\nu_1$ is better than $\nu_2$, or $\nu_2$ is better than $\nu_1$ in the given code context. Both $\nu_1$ and $\nu_2$ should be acceptable in the current code context i.e., $\exists_{i} ~score_{p_i}(\nu_1) \neq score_{p_i}(\nu_2)$.
\end{definition1}

\paragraph*{Variants are not examples}Unlike \comment{variants, clones and simions}other types, code examples need not be always similar in behavior. For instance, examples could be instructive to explain API usage in a variety of functionally different snippets.  

\paragraph*{Variants are not bug-fixes or enhancements}Let us assume $\nu_2$ is an enhancement sought over $\nu_1$. Even though an enhancement may add some desired property to the existing code, we observe that the intent has changed. $\nu_1$ and $\nu_2$ would have been variants in the earlier code context when $\nu_1$ was under development; however, in the new context, $\nu_2$ alone is acceptable and $\nu_1$ does not fit. Same applies to bug-fixes as well. A buggy-snippet does not meet the expectations of the code context, and hence is no more a candidate for being a variant. 

\paragraph*{Variants are not mutants} Mutants are artificially changed code used to assess the quality of test cases. They do not fit into the given code context. Hence, mutants are also not code variants.
\\[6pt]
In summary, similarities and differences exist among semantic clones, idioms, simions, and code examples, in terms of syntactic and semantic distance, purpose, properties, origin and connotation. Due to lack of clear definitions, the terms semantic clones, examples, and variants are often used interchangeably or even incorrectly in literature and developer discussions. In Table~\ref{tbl:CEVcomparision}, we summarize this discussion by defining, \comment{highlighting the differences}differentiating, and exemplifying each of these terms. Note that none of these types are defined or discussed along with code context or desired properties. Thus, code variants are  different from all of these types.

\subsection{Types of Variants}

\comment{Properties of Variants}
From the perspective of mining variants, developers need to decide if a particular variant is better or worse or even incomparable to the rest. To aid this activity, we introduce two categories of variants, namely, simple and complex variants.

\begin{definition1}
We refer to a set of variants as \textbf{simple} if for any pair of variants $(\nu_1, \nu_2)$ in the set, one variant scores not less than the other ($\forall_{i} ~score_{p_i}(\nu_1) \geq score_{p_i}(\nu_2)$ or $\forall_{i} ~score_{p_i}(\nu_2) \geq score_{p_i}(\nu_1)$) for all desired properties in a specific code context. Also, recall that $\exists_i ~score_{p_i}(\nu_1) \neq score_{p_i}(\nu_2)$ if $\nu_1$ and $\nu_2$ are variants. 
\end{definition1}

\begin{figure}[t]
\centering
\vspace*{-5px}
\includegraphics[trim=0cm 11.1cm 10cm 0cm, clip,width=8.5cm]{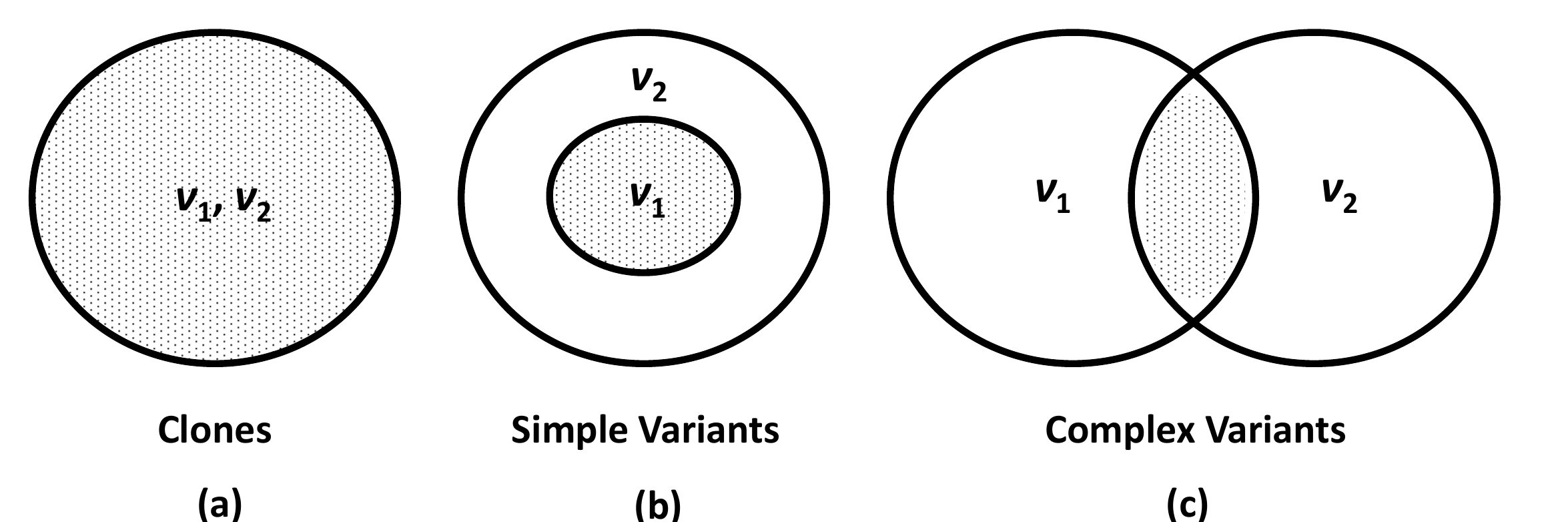}
\caption{In a given code context, if two code snippets $\nu_1$ and $\nu_2$ show the same desired properties, $\nu_1$ and $\nu_2$ are clones as shown in (a). If $\nu_2$ has even more desired properties while having all desired properties as in $\nu_1$, as shown in (b), $\nu_2$ is a stronger variant of $\nu_1$. Conversely, $\nu_1$ is a weaker variant of $\nu_2$. As shown in (c), $\nu_1$ and $\nu_2$ are complex variants if they only have a few desired properties in common. }
\label{fig:definitions}
\vspace*{-10px}
\end{figure}

If $\nu_1$ and $\nu_2$ are simple variants in the given code context, and $\nu_2$ is stronger than $\nu_1$, we mean that $\nu_2$ scores over $\nu_1$ on all desired properties (Figure~\ref{fig:definitions} (b)). Recall that by the definition of variants, there is at least one desired property by which $\nu_1$ is better than $\nu_2$. In practice, we may find that most efficient solutions may suffer from issues such as readability and licensing, and hence may not be better on all desired properties. It is possible that $\nu_1$ and $\nu_2$ may be equal in the number of desired properties they satisfy but still are not clones because the sets of properties each of them satisfy may have differences. 

\paragraph*{Strict Partial Order} \comment{The relation over the strength of code variants (represented by the symbol $\lq>\rq$) is inherently irreflexive, transitive, and antisymmetric. Let $\nu_1$, $\nu_2$ and $\nu_3$ be variants of a given code snippet. If $\nu_1$ is a stronger variant of $\nu_2$, and $\nu_2$ is a stronger variant of $\nu_3$, then $\nu_1$ is a stronger variant of $\nu_3$. We use the notation $>$ to show the stronger variant relation. Therefore, in the above case, $\nu_3$ $>$ $\nu_2$ $>$ $\nu_1$ implies $\nu_3$ $>$ $\nu_1$ (transitive). It is not possible that $\nu_2$ $<$ $\nu_1$ and $\nu_1$ $<$ $\nu_2$ by definition (asymmetric). For reflexivity, $\nu_1$ $<$ $\nu_1$ which does not hold by definition (irreflexive). There needs to be at least one property by which the two snippets being compared should differ to be called as a variant.  Thus we have a strict partial order with the variant relation `$<$'.}The relation over the strength of code variants (represented by the symbol `$>$') is a strict partial order over the set $\mathcal{V}$ of variants. In other words, $\nu_1 > \nu_1$ cannot hold (irreflexivity) since we need at least one property by which the snippet being compared with should differ to be called as a variant. $\nu_2 > \nu_1$ indicates that $\nu_2$ is a stronger variant of $\nu_1$, and $\nu_1 > \nu_2$ cannot hold (antisymmetry), and $\nu_2 \ne \nu_1$ (irreflexivity). In addition, if we have $\nu_3$ such that $\nu_3 > \nu_2$, then $\nu_3 > \nu_1$ (transitivity).

\begin{table}[t]
\small
\centering
\caption{The term ``variants" is often used by developers. For the Gabel's dataset, we explored the bug repository and found several discussions around variants.}
\vspace*{-5px}
\label{tbl:variants}
\begin{tabular}{lll}

\toprule
\textbf{System}       & \textbf{Size (in MLoc)} & \textbf{``Variants" usage} \\
\midrule
GIMP         & .78         & 109               \\
GTK          & .88         & 324               \\
MySQL        & 1.13        & 166               \\
Postgresql   & .74         & 262               \\
Linux Kernel & 7.31        & 286              \\
\bottomrule
\end{tabular}
\vspace*{-10px}
\end{table}

In the case of complex variants, it might be possible for developers to apply a weight function to choose a specific complex variant as a strong variant. Without weights or additional such preference information, it will be unclear to developers which variant to select  (Figure~\ref{fig:definitions} (c)). Figure~\ref{fig:definitions} (a) shows the case where $\nu_1$ and $\nu_2$ have the same properties, no more or no less and thus they become clones.

\begin{definition1}
We call $\nu_1$ and $\nu_2$ as \textbf{complex} variants if $\nu_1$ scores over $\nu_2$ for some desired properties, and $\nu_2$ scores over $\nu_1$ for some other desired properties. More formally, $\exists_{i,j}$ ~$(score_{p_i}(\nu_1) > score_{p_i}(\nu_2))$ \allowbreak ~$\land$~$(score_{p_j}(\nu_1) < score_{p_j}(\nu_2))$\allowbreak~$\land$~$(i \neq j)$.
\end{definition1}

\comment{example} As an example, consider internet traffic monitoring APIs such as Fiddler and Titanium. A discussion on SO\footnote{https://tinyurl.com/gr66vje} \ignore{real url is http://stackoverflow.com/questions/30995808/please-suggest-an-alternative-of-fiddler-core-3rd-party-library} suggests that Titanium is preferred over Fiddler given the licensing constraints. Hence Titanium is a stronger variant than Fiddler. In this case, Titanium is also a simple variant of Fiddler since it is easier to choose the prior over the latter. As another example for simple variant, an $O(nlogn)$ solution is accepted to be better than $O(n^2)$ solution in the context where worst-case time complexity is the desired property. Consider the use of Multinomial\footnote{https://tinyurl.com/gulbmb8}\ignore{http://grokbase.com/t/pig/dev/1353zedn9e/function-to-compute-product-of-values-in-bag} and Bernoulli\footnote{https://issues.apache.org/jira/browse/SPARK-489, \newline https://issues.apache.org/jira/browse/OPENNLP-777} Naive Bayes implementation choices. There are different contexts in which either of them is preferred. Thus they become complex variants in the context where we need additional information to categorically select one over the other. The three snippets in Figure~\ref{fig:RankingComparison}~(a) are examples for clones with respect to speed of execution since all are recursive in nature and have no further complexity.

\comment{Existing tools for variant mining} \ignore{Moreover, Juergens et al.~\cite{Juergens:2010:CSB:1955601.1955971} point out that clone detection tools find less than 1\% of total redundancies that do not originate because of copy-paste activity. Therefore, we need to understand code variants better and focus on tool support for developer productivity.} To the best of our knowledge, none of the existing tools focus on differences in desired properties for the given code context. In this sense, they are not variant mining tools. Next, we discuss the types of code contexts and desired properties in the following sub-sections.

\input{51_ExecutionContext}

\input{52_DesirableProperties}


%% file: 51_ExecutionContext.tex
\subsection{Code Context}
\label{ExecutionContext}
\comment{What do we do in this work?}While desired properties distinguish variants, the code context relates them together. Code context acts like candidate variant filtering criteria. The code context captures different aspects such as users' concerns, and development and execution contexts. As defined in Section~\ref{Definitions}, code context description comprises of one or more of the following: intent, dependencies, input/output and state. We arrived at this taxonomy based on a literature survey of 11 related research papers~\cite{Leveson,Nguyen:2012:GPC:2337223.2337232,Ponzanelli:2014:MST:2597073.2597077,Robillard:2008:TAS:13487689.13487691,Stolee:2014:SSS:2628068.2581377,Nix:1985:EE:4472.4476, Singh:2012:LSS:2212351.2212356,Rinard:2012:EPS:2240236.2240259,Bergan:2014:SEM:2660193.2660200,Dhar:2015:CSP:2786805.2786877, Buisson:2010:RES:1932681.1863550} and our experiments on nine open source projects (See Section~\ref{experiments}).



\comment{
Explain each item.
  - What does this mean? What is the scope? 
  - Give an example pair of variants in this context.
  - How to extract this context item? Why is extracting it difficult?
  - Summarize the work done in this area. What is the state-of-the-art?
  - Are there any limitations?
}

\paragraph*{Intent}Programs are products of human desire to solve specific problems or accomplish well-defined tasks. Hence, an understanding of the problem being solved plays an important part in identifying or recommending variants that ``fit" the purpose. Purpose includes functional and para-functional requirements. For example, ``computing factorial", ``implementing little endian algorithm" are examples of intent. Intent specification is a hard problem~\cite{Leveson} which goes beyond just naming and describing the problem using natural language phrases. Nguyen et al.~\cite{Nguyen:2012:GPC:2337223.2337232} relate execution context and intent. Their thesis is that the intent can be captured using the API usage patterns in the current code. Ponzanelli et al.~\cite{Ponzanelli:2014:MST:2597073.2597077} show that fully qualified name of the code elements, current code, custom API types, and method names are strong pointers to intent. 

\paragraph*{Dependencies}Often, implementation is constrained to a specific programming language, certain pre-built libraries, or components. Search for variants must honor these constraints. Constraints may also include structural elements such as methods or classes as in Java. We refer to such constraints as dependencies. For example, a REST API for financial data may be provided by multiple providers which become variant choices. In this context, we assume that non-REST APIs are not sought by the developers. Robillard~\cite{Robillard:2008:TAS:13487689.13487691} claim that neglecting such dependencies may lead to \textit{low-quality modifications}. They discuss structural dependencies in the scope of \textit{program elements} and mention \textit{methods} and \textit{fields} as examples.

\paragraph*{Input/Output}Input and output examples are used as context to search~\cite{Stolee:2014:SSS:2628068.2581377} and synthesize~\cite{Nix:1985:EE:4472.4476} source code. Nix argues that the problem of synthesizing expressions mapping given a set of inputs to the given set of outputs (in the sub-context of repetitive text editing) is NP-Hard. Programming-by-Example community shows steps taken in this direction with string transformation~\cite{Singh:2012:LSS:2212351.2212356}. In summary, a decade old research in this area has produced solutions for text editing and spreadsheet processing; however, synthesizing large sized programs remains a challenge~\cite{Rinard:2012:EPS:2240236.2240259}.

\paragraph*{State} Often, developers complain of a specific state that the system gets into. For example, in Zope\footnote{https://bugs.launchpad.net/zodb/+bug/143274}, a developer states, \textit{``For huge transactions ZEO spends a long time (in the order of minutes)
in the call to ``vote". This makes it irresponsive for other request..."}. Current context of the code under execution includes the snapshot of its variables, the line under execution, and the resources available at that time for the program~\cite{Bergan:2014:SEM:2660193.2660200}. This definition of context is used heavily in debugging~\cite{Bergan:2014:SEM:2660193.2660200}, program repair~\cite{Dhar:2015:CSP:2786805.2786877}, and real-time updates~\cite{Buisson:2010:RES:1932681.1863550} to software systems. These systems use a variety of techniques such as automata and logic for capturing and representing the context.



%% file: 52_DesirableProperties.tex
\subsection{Desired Properties}
\label{sec:desiredproperties}
\begin{table*}[!t]
\small
\caption{Examples of developer discussions taken from open source projects describing the desired properties (of different types) in code variants.}
\label{PropertyExamples}
\vspace*{-10px}
\begin{tabular}{p{1.6cm} | p{2cm} | p{7.2cm}}
\toprule
\textbf{Type} & \textbf{Issue} & \textbf{Developer Discussion} \\
\midrule
Algorithmic & Eclipse Bug 384730
 & \textit{There are already some implementation of this algorithm. However, most of them are pretty complex and slow. I would like to contribute a smaller and simpler version compatible with the ZEST layout engine.} \\
Algorithmic & Eclipse Bug 409427
 & \textit{We should replace the existing xml format with the much denser and faster loading EMF BinaryResource implementation.} \\
Resource- oriented & HTTPClient-1267 & \textit{Does httpcomponents not support transient cookies? If so we cannot use this library for session login, since most session-protected sites use this.} \\
Resource- oriented & Eclipse Bug 293637 & \textit{Ribbon must be licensed by each adopter. If Eclipse will provide Ribbon, than every RCP application with Ribbon must be licensed. This violates EPL.} \\
Diction & Eclipse Bug 196585
& \textit{It is better to use the setter methods on the model classes (e.g. TracWikiPageVersion) than having constructors with many parameters. That way the order of the parameters does not get mixed up and the code is easier to refactor and to read.} \\
Diction & Eclipse Bug 338065 & \textit{Our coding conventions currently demand to declare all method parameters as final in order to prevent parameter assignments. Meanwhile, parameter assignments can effectively be revealed by the Eclipse tooling and by tools like FindBugs on the CI server.} \\
\bottomrule
\end{tabular}
\end{table*}

\comment{Classifying desired properties}As discussed in Definition~\ref{def:dprops}, desired properties distinguish variants. Desired properties in a variant can be classified into broadly three groups: a) Algorithmic, b) Resource-Oriented and c) Diction. 

\paragraph*{Algorithmic Properties}Algorithms play a significant part in computation and their properties are well studied~\cite{Cormen:2009:IAT:1614191}. Developers seek efficient algorithms to make their code score on para-functional attributes such as security~\cite{Shahriar:2012:MPS:2187671.2187673}, accuracy~\cite{Jiang:2007:DSA:1248820.1248843}, readability~\cite{Moreno:2015:IUT:2818754.2818860} and scalability~\cite{Gabel:2008:SDS:1368088.1368132}. Sridhara et al.~\cite{Sridhara:2011:ADD:1985793.1985808} discuss the importance of identifying high-level algorithmic steps in source code. Patterns, signatures and structures are limited in their ability to detect algorithms in source code~\cite{Mishne:2004:SCR:2816272.2816322}. Many reuse techniques~\cite{Bajracharya:2010:LUS:1882291.1882316,Wang:2014:ACS:2642937.2642947} that work at function level or code fragment level focus on semantic similarity and ignore the variability across variants. Mishne et al.~\cite{Mishne:2004:SCR:2816272.2816322} extract concept graphs to represent algorithmic information. They use a finite list of concept types, such as loop, assign and block, in their representation. Yet, this model shows good results for code based on C language. 

\paragraph*{Resource-oriented Properties}For reasons such as licensing~\cite{Wu:2015:MDL:2820518.2820558,Alspaugh:2009:ASL:1572192.1572203}, certain libraries, components, sub-systems, interfaces, and services are considered better or relevant. This property has nothing to do with the syntax or semantics of the code snippet. Instead, it is about the extraneous (non-code) elements associated with the snippet, such as the legal constraints, and trust factors. Long~\cite{Long:2001:SRA:505482.505492} observes that many third-party libraries are no longer actively maintained. He calls this the \textit{used car fiasco}. He brings up more issues in reuse, such as \textit{One size fits all} and \textit{Of course it's reusable}. Moreno et al.~\cite{Moreno:2015:IUT:2818754.2818860} discuss the effort to reuse the code snippet.

\paragraph*{Diction Properties} Diction refers to the \textit{style of speaking or writing as dependent upon choice of words}\footnote{http://www.dictionary.com/browse/diction}. Some developers may prefer {\codefont{for}} over {\codefont{while}} to code a loop. Resulting code is semantically the same. Naming conventions may contribute to the ranking of one variant over the other~\cite{Allamanis:2015:SAM:2786805.2786849}. We call such variants as diction variants. Diction variants cover all non-algorithmic and non-resource-oriented properties, such as patterns, refactoring needs, conventions and style. Often, programming language libraries give multiple ways to implement the same functionality within the same resource and algorithmic constraints. Syntactic sugars~\cite{Pombrio:2014:RLE:2666356.2594319} are classic examples for this type of variants. 

\ignore{
\begin{table}[!t]
\small
\centering
\caption{A diction variant for the non-recursive code shown in Table~\ref{AlgorithmicPropertyExample}. Note that a few variable names are different. Otherwise, the code is algorithmically the same and depend on the same kind of resources.}
\label{DictionPropertyExample}
\begin{tabular}{p{8cm}}
\vspace{-5pt}
\toprule
\begin{lstlisting} 
String reverse(String str) {   
    StringBuilder temp = new StringBuilder();
    for (int i = str.length() - 1; i >= 0; i--)
        temp.append(str.charAt(i));
    return temp.toString();
}
\end{lstlisting}
\bottomrule
\end{tabular}
\vspace{-10pt}
\end{table}
}
Certain syntactic choices have distinguished benefit over the other. A recursive version of factorial is rarely used in practical scenarios. A memoization approach avoids recomputation and is desired especially when large inputs values are bounded so that a four byte variable such as Java \textit{int} can hold the result. Yet, the role of diction variants have been largely ignored by the research community. In academic context, most plagiarism tools depend on these differences to avoid marking student works as duplicate. 


\comment{Why should we worry about these properties?}Absence of one or more of these properties leads to low-quality code snippet for which developers seek replacement. This absence may introduce faults, bad smells or sub-optimal code. Table~\ref{PropertyExamples} shows real developer discussions from Eclipse and HTTPClient projects. We have mapped these discussions to one of the three types of properties discussed. 



%% file: 6_experiments.tex
\section{Empirical Evaluation}
\label{experiments}
Here we present the results of our experimental studies structured around a few Research Questions (RQ). The experiment data is shared at the project website\footnote{\url{http://variants.usite.pro/index.html}}\ignore{\footnote{http://tools.pag.iiitd.edu.in:8092/variants/index.jsp}}.

We use a dataset of fifteen open source projects (Table~\ref{tbl:listofprojects}). When making the project selection, we wanted to accommodate a variety of programming languages, domains of use, and project sizes. Three of the projects in our dataset are written in Java, three in C, and the remaining three are in Python. Java is a statically typed and compiled object-oriented language whereas C is not object-oriented, but is static (weakly) typed. Python is a dynamically (strongly) typed and interpreted object-oriented language.  

Allamanis et al.~\cite{Allamanis:2013:MSC:2487085.2487127} evaluated idioms on popular Java projects that were selected based on their z-score in GitHub. We pick Atmosphere and Hibernate from that list. To ensure a variety in the project domains, we replaced one project from the list of Java projects used by Allamanis et al.,~\cite{Allamanis:2013:MSC:2487085.2487127} with Apache Math which is an algorithmically sensitive library. Three of these projects, GIMP, GTK+ and MySQL, written in C, were used by Gabel et al.~\cite{Gabel:2008:SDS:1368088.1368132} in their work on semantic clones. Roy et al.~\cite{Roy:2010:SLR:1808901.1808904} in their study of code clones used the Python projects: Plone, SCons, and Zope, which we also select. 
The projects in our dataset vary from 68K (Atmosphere) to 1130K (MySQL) lines of code, and include disparate domains such as mathematics, databases and editors. 


\subsection*{RQ 1: Do variants matter?}
\label{DoVariantsMatter} Before exploring different properties of variants in depth, we investigated about the relevance of variants and about their nature of presence in the development community. Here we seek to understand how often developers make choices about variants in their projects. 

\begin{table}[]
\small
\centering
\caption{Projects dataset used to evaluate if variants exist in real open source projects. Diversity in programming languages (Java, C and Python), size and domain characterizes these projects.}
\label{tbl:listofprojects}
\vspace*{-5px}
\begin{tabular}{l|p{1.8cm}|p{1cm}|l|p{3cm}}
\toprule
\# & \textbf{Project} & \textbf{PL} & \textbf{LoC} & \textbf{Domain}           \\
\midrule
1  & Apache Math      & Java              & 375K         & Mathematics                      \\
2  & Atmosphere       & Java              & 68K          & Client-Server   \\
3  & Hibernate ORM    & Java              & 930K         & Domain Model persistence  \\
4  & Gimp             & C                 & 780K         & Image Manipulation        \\
5  & GTK{\plus}             & C                 & 880K         & UI Widget Toolkit \\
6  & MySQL            & C                 & 1130K        & Database                        \\
7  & Plone            & Python            & 74K          & Content Management \\
8  & SCons             & Python            & 228K          & Build Tool \\
9  & Zope             & Python            & 272K         & Web Application Server \\ \bottomrule  
\end{tabular}
\vspace*{-3px}
\end{table}


We took 100 random defects in each of the fifteen projects in our dataset. When we analyzed the discussion in these defects, we found 277 variants as shown in Table~\ref{tbl:dovarmatter}. This indicates that variants exist in abundance. Developers actively compare and seek variants. Table~\ref{tbl:varintsinso} shows a sample of such discussions. Discussions were categorized as about variants if they discussed different forms of implementation or were about implementation choices.
We found that variant discussions exist in \textbf{25\%} (in Zope) to \textbf{43\%} (in Apache Math) of bug reports in these projects. \remark{AS:(people actively compare and seek variants, and simple variants feature in discussion since it is patch/code review process)}Discussions about simple variants occurred 17\% times on average, as compared to 14\% for complex variants. Discussions about simple variants were uncomplicated, where a developer proposed a variant in a patch and was approved. Discussions about complex variants were more involved, going over the pros and cons with respect to the desired properties of the variants.  

Overall, \emph{31\% bugs discussed variants}. Hence, we conclude that variants play a significant role in software development. This serves as a strong motivation to further explore variant characteristics. We discuss these in the following RQs. 

\begin{table}[!t]
\small
\centering
\caption{Volume of variant discussions in open source projects depends on the project domain. $Alg_v$, $RO_v$ and $D_v$ represent algorithmic, resource-oriented and diction variants respectively. Dominating types are in bold. }
\label{tbl:dovarmatter}
\begin{tabular}{p{1.8cm}|r|r|r|r|r|r|r}
\toprule
\multirow{2}{*}{\textbf{Project}} & \multicolumn{3}{c|}{ \textbf{Simple} } & \multicolumn{3}{c|} {\textbf{Complex}} & \multirow{2}{*}{\textbf{Sum}} \\ \cline{2-7}
 & \textbf{$Alg_v$} & \textbf{${RO_v}$} & \textbf{{$D_v$}} & \textbf{$Alg_v$} & \textbf{${RO_v}$} & \textbf{{$D_v$}} \\
\midrule
ApacheMath & 4 & 1 & 8 & 11 & 7 & \textbf{12} & 43 \\
Atmosphere & 2 & 7 & \textbf{8} & 6 & 2 & 5 & 30  \\
Hibernate & 5 & 3 & \textbf{10} & 2 & 4 & 5 & 29 \\
GIMP  & \textbf{11} & 2 & 9 & 7 & 3 & 3 & 35   \\
GTK{\plus}  &  6 & 3 & 7 & \textbf{9} & 3 & 4 & 32\\
MySQL & \textbf{11} & 1 & 8 & 6 & 4 & 1 & 31 \\
Plone & 2 & 2 & \textbf{10} & 4 & 3 & 5 & 26 \\
SCons & \textbf{8} & 0 & 6 & 3 & 2 & 7 & 26 \\
Zope & \textbf{10} & 1 & 2 & 6 & 3 & 3 & 25 \\
\midrule
\textbf{Average} & 7 & 2 & 8 & 6 & 3 & 5 & 31 \\   
\bottomrule
\end{tabular}
\vspace*{0px}
\end{table}

\begin{table}[!t]
\small
\centering
\caption{Developer discussions on implementation choices is common in SO. Some questions from SO in which multiple implementation choices were discussed are given below along with related programming language and the reference ID to look up the discussion item.}
\label{tbl:varintsinso}
\begin{tabular}{l|p{1cm}|p{5.5cm}}
\toprule
\textbf{Ref} & \textbf{Tag} & \textbf{Question}\\
\midrule
7519283 & sql & Fastest way to check if a character is a digit?\\
1102891 & java & How to check if a String is numeric in Java   \\
53513 & python & Best way to check if a list is empty     \\
392022 & bash & Best way to kill all child processes
\\ \bottomrule
\end{tabular}
\vspace*{-5px}
\end{table}

\ignore{
\begin{table}[!t]
\small
\centering
\caption{Same code snippet pair can be variants of different types.}
\label{tbl:multi-type-variants}
\begin{tabular}{l|p{2cm}|p{2cm}|p{2cm}}
\toprule
\textbf{Intent} & \textbf{Variant type}      & \textbf{Project}             & \textbf{\#Bug}   \\
\midrule
Sine   & Resource-Oriented & Apache Commons Math & MATH-901 \\
Sine   & Algorithmic        & Apache Commons Math & MATH-901           
\\ \bottomrule
\end{tabular}
\vspace*{-2px}
\end{table}
}

\subsection*{RQ 2: How are variants distributed across the types of desired properties?}
\ignore{We investigated this in two steps. In step one, we explored SO questions to collect topics and their corresponding variants. In step two, we looked up these variants in open source projects. SO helped us to understand variant implementations for several topics. Table~\ref{tbl:varintsinso} shows a sample of such discussions. For example, ``reversing a string" is discussed in several posts, such as ``\textit{Reverse a string in Java}", ``\textit{How do you reverse a string in place in JavaScript?}", ``\textit{c\# - Best way to reverse a string}", ``\textit{java - Easy way to reverse String}", and ``\textit{php - Reverse string without strrev}". For this intent, we then searched for distinct implementations in issue trackers and source code of open-source projects. We found three algorithmic variants and four diction variants (Table~\ref{tbl:reverse}). Similarly, we analyzed the SO discussions, the code or library referenced in the post, and used it to (web) search for their implementations in real world projects (Table~\ref{tbl:desiredprops}). We searched through SO posts until we had a set of 100 implementation choices. These variants covered 34 intents ranging between two to seven implementation choices per intent (averaging three), and were used in 88 open source projects, across 13 different programming languages. } 
As developers are constantly looking for variants to reuse, they may look out for certain properties that make one variant preferred over others for the given project requirements. We were interested in knowing the distribution of these variants across different properties. We looked up the discussions talking of variants in the fifteen open source projects to analyze the distribution of these variants with respect to their types: algorithmic, resource-oriented, and diction. 

Algorithmic variants were typically discussed because new algorithms were sought in new requirements, enhancements or bug reports. Table~\ref{tbl:dovarmatter} shows that there were on average 7\% simple and 6\% complex algorithmic variants. Thus, on an average, 13\% algorithmic variants were found in these bug reports. Only 5\% variants (2\% simple and 3\% complex) belonged to the resource-oriented class. They are discussed only when there is a concern or conflict such as licensing, library compatibility and coding conventions. These concerns can be seen in tracking bugs which call for changes at multiple places while keeping the bug count to just one. In two of the projects (MySQL and Apache Math), we found only one resource-oriented variant in the 100 randomly chosen defects. MySQL had only one simple resource-oriented variant. This does not mean that there are very few resource-oriented variants in these projects. It only suggests that the distribution is skewed towards other types of variants. Moreover, constraints such as licensing, library compatibility and coding conventions are expected to be known to seasoned developers, which could be another reason for a relatively lower count of resource-oriented variants discussions when compared with the count based on other properties discussed. A time-boxed, focused search for an hour resulted in 20 resource-oriented MySQL variants, which shows that discussions on these resource-oriented variants exist, but are less frequent. On average, 13\% of discussions were about diction variants (8\% simple and 5\% complex). This makes a case for the use of type-1 to type-3 clone detection tools as they are useful in this space (diction variants).

Drawing such characteristics could be useful for developers in mining variants based on the properties best suited for the project requirements. For instance, developers looking at optimising program function involving data clustering may be looking for algorithmic variants of clustering. This led us to RQ 4 which studies if these desired properties among variants are dependent on the domain of the project.

\ignore{
\begin{table}[!t]
\small
\centering
\caption{Variants spotted for the topic ``Reverse" in open-source projects.}
\label{tbl:reverse}
\begin{tabular}{l|l|l|p{2cm}}
\toprule
\textbf{Variant type} & \textbf{Project}      & \textbf{Language} & \textbf{\#Bug or file}                 \\
\midrule
Algorithmic   & Ruby         & C        & 12744                         \\
Algorithmic   & CloverETL    & C        & CLO-1153                      \\
Algorithmic   & Bro          & C++      & BIT-969                       \\
Diction    & myconnector  & Java     & TestService-Impl.java               \\
Diction    & netbeans-soa & Java     & GenUtil.java                  \\
Diction    & elexis       & Java     & Directories-Content-Parser.java \\
Diction    & codelab      & Java     & JavaBasics.java              
\\ \bottomrule
\end{tabular}
\vspace{-5px}
\end{table}
}

\begin{table}[t]
\small
\centering
\caption{Variants are often sought and their desired properties are discussed in bug reports of real-world projects.}
\label{tbl:desiredprops}
\small
\vspace*{-5px}
\begin{tabular}{p{1.4cm}|p{1.5cm}|p{2cm}|p{1.5cm}|p{2cm}}
\toprule
\textbf{Project}             & \textbf{Intent}   & \textbf{Variant type }     & \textbf{Language}   & \textbf{\#Bug} \\ 
\midrule
Ruby                & Reverse  & Algorithmic        & C          & 12744     \\
Apache Commons Math & Sine     & Resource-Oriented & Java       & MATH-901  \\
CiviCRM             & Maps API & Resource-Oriented & PHP        & CRM-524   \\
Moodle              & charting & Resource-Oriented & Javascript & MDL-55090 \\
ReactOS             & Convert  & Resource-Oriented & C          & CORE-5849 \\
Apache Lens         & Query    & Algorithmic        & SQL        & LENS-1028 \\  
Hadoop Map Reduce & Get NameNode URI & Diction & Java & MAP-REDUCE-6483
\\ \bottomrule
\end{tabular}
\vspace*{-5px}
\end{table}

Further, we found that 41\% of the intents lead to algorithmic variants (Table~\ref{tbl:dovarmatter}). These were largely due to differences in requirements of performance or accuracy of results. \ignore{Table~\ref{AlgorithmicPropertyExample} gives an example from SO\footnote{http://stackoverflow.com/questions/15996885/which-variants-string-reverse-are-better} where the developers discuss reversing a  string in Java.}
For example, in Ruby project, an implementation referred to as {\codefont{reverse.each\_char}} that uses {\codefont{rb\_enc\_left\_char\_head}} covers all encodings to scan a string backwards~\footnote{https://bugs.ruby-lang.org/issues/12744}. Overall, 18\% of intents led to resource-oriented variants. These were largely due to limitations and constraints (such as license) on the reuse of third-party libraries. For example, porting existing code from a different programming language was cited as an issue in some of the bug reports. Finally, diction variants were also a significant part of developer discussions. About 41\% of existing variants belonged to this type. Typical diction variants discussions were about better coding conventions and styles.

\begin{table}[t]
\small
\centering
\caption{Code contexts are primarily described using one or more of these four types. We map bug descriptions containing variants to these types. \ignore{Complete list for all fifteen projects is on the website.}}
\label{ECTypes}
\vspace*{-5px}
\begin{tabular}{l|l|l|l|l|l}
\toprule
\# & \textbf{Project} & \textbf{Intent} & \textbf{Dep.} & \textbf{I/O} & \textbf{State} \\
\midrule
1 & Apache Math (Java) & 33 & 11 & 2 & 0  \\
2 & Atmosphere (Java)      & 24              & 12                  & 2                     & 1              \\
3 & Hibernate ORM (Java)& 18 & 9 & 1 & 3 \\
4  & Gimp (C)             & 16              & 6                   & 3                     & 10             \\
5 & GTK+ (C)& 23 & 4 & 4 & 8 \\
6 & MySQL (C)& 21 & 3 & 9 & 9 \\
7  & Plone (Python)           & 19              & 7                   & 8                     & 6             \\
8 & SCons (Python) & 17 & 10 & 0 & 3 \\
9 & Zope (Python) & 20 & 7 & 4 & 10 \\
\bottomrule
\end{tabular}
\vspace*{-5px}
\end{table}

\begin{table*}[!th]

\small
\centering

\caption{Variant discussions from the dataset capturing the code contexts. }
\vspace*{-5px}
\label{ECSamples}
\begin{tabularx}{1\textwidth}{p{1.55cm}|p{1.7cm}|X|p{2.8cm}}
\toprule
\textbf{Code Context} & \textbf{Bug ID} & \textbf{Excerpts from the Discussion} & \textbf{Essence of the Discussion w.r.t the context type} \\
\midrule
Intent & MATH-785 &  \textbf{S1:}The ContinuedFraction calculation can underflow in the evaluate method, similar to the overflow case already dealt with. \textbf{S2:}The evaluation of the continued fraction has been changed to the modified Lentz-Thompson algorithm which does not suffer from underflow/overflow problems as the original implementation. & Variants have a common intent here which is to compute continued fraction. \\
State & Atmosphere-2037     & \textit{\textbf{S1:}the \_disconnect method is not being called for Android Chrome when closing the browser though the android apps panel. Only in case of changing the URL on the browser, that method is called. \textbf{S2:}It works with Firefox desktop, Firefox mobile, chrome desktop, etc.}  &  State of disconnect method invocation on closing different browsers being discussed. \\
Input/ Output & MATH-1143 &  \textit{\textbf{S1:}A DerivativeStructure and UnivariateDifferentiableFunction are great tools if one needs to investigate the whole function but are not convenient if one just needs derivative in a given point. \textbf{S2:}Give the derivatives in the ``natural'' order, which is in increasing order when you have one parameter and high order derivatives, and in parameters order when you have only first order derivatives for all parameters.} & Common output of derivative, across all variants. \\
Dependency  & MATH-1098       &        \textit{As we will certainly not add a dependency to another library, we could start with our own set of annotations} & Two variants of a function where both do not have a dependency on another library.\\
\bottomrule
\end{tabularx}
\end{table*}

\subsection*{RQ 3: How are variants distributed across the context types?} Recall that we identified intent, input/output, dependencies and state as the four major constituents of code context. To study the characteristics based on which one could classify one code as a variant of another it is essential to spot the common attributes among variants. To find such attributes we investigate the variant discussions. Further, to validate if the identified four characteristics indeed capture the code context of variants, we classify the contexts described along with the variant discussions in our dataset into these four constituents. Table~\ref{ECSamples} gives an essence of the characteristics based on which the contexts were classified. We were successful in classifying all the contexts into these four constituents. We randomly selected 100 of these defects and cross-verified to ensure inter-rator agreement. We found no ambiguity in the understanding of context boundaries and thus the categorization was found to be accurate. Table~\ref{ECTypes} shows the breakdown per project. \remark{AS:where is this data from, refer to the table}We found that intent dominates (in 191 out of 343 contexts found i.e., 56\%) across the bug reports as most reports discussed were built upon the underlying functionality of code variants. Dependency, input/output and state covered 20\%, 10\% and 15\% of contexts in bug reports, respectively.  
Identifying the nature of similarities could help developers filter out candidates for variants.

\ignore{
\begin{figure}[t]
\centering
\includegraphics*[trim=1cm 1cm 1cm 1cm, clip,height=3cm]{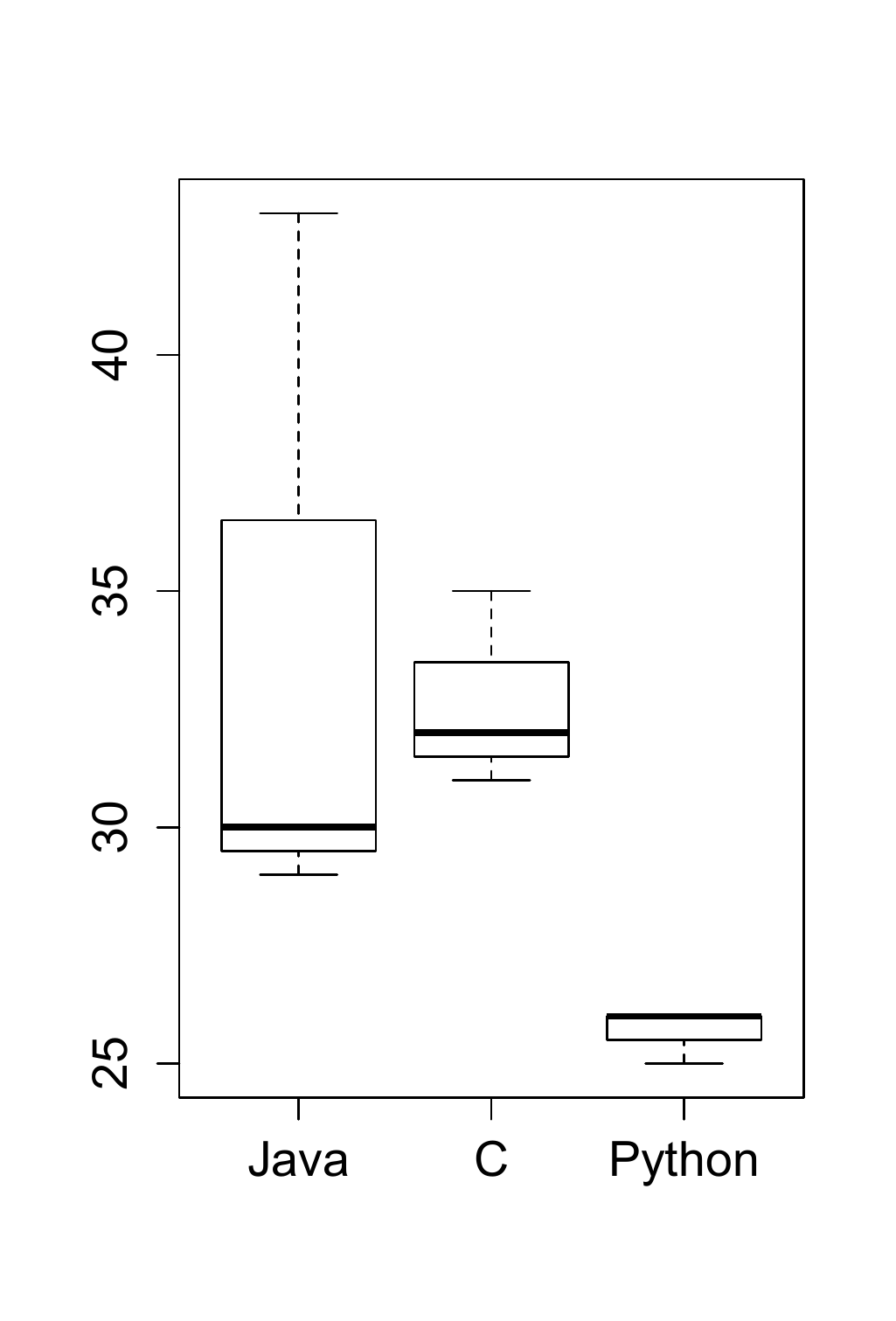}
\caption{Reuse in source code happens at different levels for different reasons. This work focuses on code variants.}
\label{fig:redundancy}
\end{figure}
}

\subsection*{RQ 4: Are variants domain-dependent?} 
We investigated whether the nature of the project (domain) is connected to the existence or importance of the variants. We observed that the distribution of the type of variants changes heavily across projects. Dominant types are highlighted (in bold) in Table~\ref{tbl:dovarmatter}. Algorithmic variants dominated in GIMP. As a project about image editing, the code snippets discussed were on intents, such as finding optimal palettes and usage of image masks that were algorithmic. Atmosphere is a client server framework, and it discussed several (30\% i.e., 9 out of 30) resource-oriented variants. Diction variants appeared in large numbers (58\% i.e., 15 out of 26) in Plone as it talked about removing deprecated exception handling syntax and labels that were not in sorted order in the code. Some sample discussions supporting such observations are shown in Table~\ref{domsample}. From these instances, we gather that domain characteristics influence reuse in different ways, such as library usage, licensing concerns, algorithmic depth, performance requirements, and so on. This leads to an uneven distribution of variants across types. 

\clearpage
\begin{longtable}{p{1.7cm}|p{1.3cm}|p{1.7cm}|p{3.3cm}|p{2cm}}
\caption{Samples from the dataset showing the influence of domain on distribution of variants.}
\label{domsample}\\
\hline
\textbf{Code Context} & \textbf{Domain} & \textbf{Bug ID} & \textbf{Excerpts from the Discussion} & \textbf{Essence of the Discussion w.r.t the domain} \\
\hline
GIMP (Algorithmic) & Image Manipulation & GIMP-66257  & \textit{ \textbf{S1:}When the Gimp converts an image to indexed mode using ``Generate Optimal Palette" and there are more colors in the RGB image than the palette can hold, then all colors in the image are changed. This is not a problem in some cases (and it is even desirable most of the time because the averaged colors give better results) but in other cases this can lead to bad results if some of the colors had to keep a precise value. \textbf{S2:}One relatively painless method/hack would be to add yet another pass over the image data which snaps all of the colours in the final palette back to their closest in-image colour before doing the final remap or dither.} & Precision being discussed across variants of converting image to indexed mode. \\
Atmosphere (Resource-Oriented) & Client-Server & Atmosphere-888     & \textit{Add timeout support for WebSocket to prevent thread waiting indefinitely}  &  A variant to better manage threads(resources). \\
Plone (Diction) & Content Management & Plone-732 &  \textit{The toolbar displays the "xx days ago" information during the loading of a page as ISO date time string and then turns it into ``xx days ago" after the complete loading of the HTML. This is an visual artifact and confuses the eye especially when the page is loaded over a slow line or takes some time for rendering. The toolbar should only display ``xx days ago" and not the ISO string at all. Perhaps the toolbar or this particular toolbar items should be hidden by default and made visible after moment.js} & Alter the content displayed. \\
\hline
\end{longtable}
\vspace*{-5px}

\subsection*{RQ 5: Are variants language dependent?}
We analyzed if programming languages affected the volume of variants (see Table~\ref{tbl:listofprojects} and~\ref{tbl:dovarmatter}). It was interesting to see if constructs of a given programming language impacted implementation choices. Automated techniques for variant mining will be harder to devise if the volume of variants are programming language dependent. 

Volume of variants in Java projects ranged from 29\% to 43\%. C projects ranged from 31\% to 35\% (See Sum column in Table~\ref{tbl:dovarmatter}). Python has the least occurrence of variants of 25\% to 26\%. Python projects were considerably smaller as compared to Java and C projects. Hence, that could be one reason for reduced discussion on variants. Overall, we do not see any conclusive evidence to show that programming languages affect the existence of variants.



%% file: 61_applications.tex
\section{Application of Variant Characterization}
As an application of the usage of characterization of variants, we investigate ranking of code search results based on ``desired properties" in which a user is interested. Search engines typically rank results based on relevance of query terms. Recently, researchers have focused on the diversity of search results based on structures and semantics of source code~\cite{Martie:2015:SEC:2820518.2820530}. Our hypothesis is that diversity of search results based on a set of desired properties, is helpful.

\paragraph*{Problem Statement} We formulate the problem of ranking code snippets for a query phrase based on the strength of desired properties.\ignore{ Let $q$ be the query phrase consisting of a set of $n$ terms \{$t_1,t_2,...,t_n$\}.} Our objective is to compute a $score_P(\phi, q)$, where $P$ denotes the set of desired properties and $q$ is a query phrase. Each property $p_i$ is represented using natural language terms for a specific code snippet $\phi$. The $score_P(\phi, q)$ gives the strength of the variants. \remark{AS: this is not making sense, get stuff from Skype} Recall that two code snippets with no difference in desired properties should have the same $score_P(\phi, q)$ as mentioned in Definition~\ref{clonedefn}. 

\paragraph*{Building a knowledgebase of snippets and their properties} We observed that SO posts contain discussions on desired properties of code snippets. \remark{AS: from where do you know these properties}Our approach started with the construction of a list of known desired properties. \ignore{This step can be automated as shown in \cite{}.} For each property, we collected stemmed synonyms $(p_i, syn(p_i))$ and antonyms $(p_i, ant(p_i))$. We used the Snowball stemmer~\cite{Stemmer} for this purpose. From each SO post, we collected the code snippets, and then computed the term frequencies ($tf(.)$) of synonyms and antonyms. After this exercise, we ended up with a knowledgebase of triples ($\phi, p_i, score$), where score (see Equation \ref{score-computation}) is the strength of desired property $p_i$ demonstrated by the code snippet $\phi$. This score can be used to boost the ranking of snippets based on desired properties while using ranking algorithms like BM25~\cite{Robertson:2004:SBE:1031171.1031181}.
\begin{equation}
\label{score-computation}
\mathit{score_{p_i}}(\phi) = \sum_{s \in syn(p_i)}tf(s) - \sum_{a \in ant(p_i)}tf(a)
\end{equation}

\paragraph*{Knowledgebase compression} We computed scores for each snippet per post. Duplicate code snippets across posts carried different scores for the same property since the terms surrounding them were different. As the frequency of snippets increase, the average of the scores converge. 

There are a variety of approaches to compute structural similarity of source code~\cite{Higo:2014:WMF:2635868.2635886,Su:2016:CRD:2950290.2950321,Schleimer:2003:WLA:872757.872770}. We used the de-duplication approach explained in \cite{Vinayakarao:2015:SHS:2678015.2682537} for its simplicity and effectiveness for method-level similarity computation. In this approach, code $\phi$ is transformed into a set of predefined terms $\psi$ to represent programming structures. This approach traverses through the code snippet, and collects loops, operators and conditionals in the same sequence as they appear in the snippet. Assignments, returns, method declarations, comments and rest of the code except loops, conditionals and expressions, are ignored. For instance, the transformed snippet for the first snippet in Figure~{\ref{fig:RankingComparison}(a) is {\codefont{if<= if* if-}}. This denotes that the operators {\codefont{<=, *}} and {\codefont{-}} are used inside a conditional statement. The structural similarity between two code snippets $\phi_1$ and $\phi_2$ is:
\begin{equation}
\label{similaritycomp}
\mathit{similarity}(\phi_1,\phi_2) = \frac{|\psi_1 \cap \psi_2|}{\mathit{max}\{|\psi_1|,|\psi_2|\}}	
\end{equation}

As we merge duplicate snippets, we aggregate their scores as {\codefont ($\phi, p_i, \sum(score)$)}. Snippets that are exactly the same will have a similarity score of 1, and snippets with no common transformed structures will have a similarity score of 0.

\begin{figure*}[!th]
\centering
\begin{subfigure}{\textwidth}
\begin{minipage}{.3\textwidth}
\begin{lstlisting}
public float factorial
	(float num){
 if(num<=1) 
   return 1;   
 else 
   return num * 
     factorial(num-1); 
}
\end{lstlisting}
\end{minipage}
\hfill
\begin{minipage}[l]{400cm}
\begin{lstlisting}
public int factorial
         (int input){
if(input==0)
	return 1;
else
	return input*
	 this.factorial
	      (input-1);
}
\end{lstlisting}
\end{minipage}
\hfill
\begin{minipage}{.2\textwidth}
\begin{lstlisting}
int fact(int n) {
	 if ( n==1) 
	 	return 1;
	 return 
	 	fact (n-1) * n;
}
\end{lstlisting}
\end{minipage}
\hfill
\vspace*{-5px}
\caption{Top-3 results from CodeExchange for query ``factorial" before applying our ranking algorithm.}
\end{subfigure}

\begin{subfigure}{.9\textwidth}
\begin{minipage}{.2\textwidth}
\begin{lstlisting}
public static double 
   factorial(int s) {
 if (s < 0 || s > 17)
   return Double.NaN;
 double[] a = { 1.0,
   1.0, 2.0, 6.0, 24.0,
   ... 	
   355687428096000.0 };
   return a[s];
}
\end{lstlisting}
\end{minipage}
\hfill
\begin{minipage}[l]{200cm}
\begin{lstlisting}
public BigInteger 
   asBigInteger()
{
  BigInteger result = 
    BigInteger.ONE;
  for (int i = 2; i <= n; 
    ++i) result = result.
	   multiply( BigInteger.
	     valueOf(i));
	return result;
}
\end{lstlisting}
\end{minipage}
\hfill
\begin{minipage}{.2\textwidth}
\begin{lstlisting}
public int factorial
            (int n) {
if (n < 0) throw new 
 CustomException();
if (n == 0)
 return 1;
if (n == 1)
 return 1;
return n * 
  factorial(n - 1);
}
\end{lstlisting}
\end{minipage}
\hfill
\caption{Top-3 results after ranking on the desired property of ``speed of execution" for the query ``factorial".}
\vspace*{-5px}
\end{subfigure}
\caption{Comparison of results before and after r-ranking.}
\label{fig:RankingComparison}
\vspace*{0px}
\end{figure*}

\paragraph*{Ranking search results} Given a set of search results, we matched each code snippet in it with the knowledgebase. We used the same Equation~\ref{similaritycomp} to find similarity of code snippets. To compute the score on multiple properties, we computed the score for each property ($p_i$) and added them to get the overall score, ($\phi, P, \sum(score)$). Figure~\ref{fig:RankingComparison} shows the top three results before and after applying our ranking algorithm. Notice that the first result is a switch-case based implementation, which is argued to be faster in SO. Apache Math uses this implementation for factorial. 

\paragraph*{Structural Heterogeneity}  Results can be improved by dropping all but the first result that look structurally similar. CodeExchange~\cite{Martie:2015:SEC:2820518.2820530}, a code search engine, shows 16 recursive factorial results out of the first 20 results. Since we are only interested in diversity with respect to desired properties, we used the approach detailed in~\cite{Vinayakarao:2015:SHS:2678015.2682537} to drop similar results at each rank. Figure~\ref{fig:RankingComparison} shows the results after application of structural heterogeneity.

\paragraph*{Evaluation} To evaluate this approach, we took the top-10 results of a search on ``factorial" in CodeExchange, and compared it with results from our ranking approach for the desired property, ``speed of execution", over the same underlying content. We chose speed of execution as it plays an important role in software development, especially in mobile environments~\cite{Liu:2014:CDP:2568225.2568229}. We took the top-10 results after applying our ranking algorithm on the same content as CodeExchange (code snippets from GitHub), for each query and compared them with the top-10 results of CodeExchange. To decide if a code snippet is a variant, we used Definition~\ref{clonedefn} and Definition~\ref{variantdefn}. We compared the search results ranking using Mean Average Precision (MAP) over ten queries as shown in Table~\ref{tblRerank}. Our approach improves the MAP from 0.17 to 0.51. 

\begin{table}[]
\small
\centering
\caption{Average precision before ($P_1$) and after ($P_2$) the application of re-ranker for 10 queries for the desired attribute of ``speed of execution".}
\label{tblRerank}
\vspace*{-5px}
\begin{tabular}{l|l|l|l|l|l}
\toprule
Query                 & $P_1$ & $P_2$ & Query & $P_1$ & $P_2$ \\
\midrule
factorial             &     0.1      & 0.7 &read from file &  0.3 &  0.6   \\
substring search      &     0.1 & 0.6       &         parse xml  & 0.2 & 0.6     \\
finding duplicate     &      0.2 & 0.5      &         calculate mean &  0.1 & 0.4    \\
matrix multiplication &    0.1 & 0.2        &         iterate over list & 0.1 & 0.4    \\
serialization         &    0.1 & 0.6        &         write log file &   0.4 & 0.5   \\
\bottomrule
\end{tabular}
\vspace*{-5pt}
\end{table}


\paragraph*{Discussion} Our approach is a proof of concept for mining variants. Precision of our approach depends on the precision of several sub-steps, such as similarity computation, de-duplication, synonyms and antonyms computation. We assume that if a desired property is strongly demonstrated in a code snippet, that property will be discussed in the SO post. This assumption may not hold true in all posts. Also, the synonyms and antonyms that show up in the post may refer to code snippets other than the one added to the answer. Further, we have ignored negative qualifiers, and hence our results are not always accurate. For example, ``not efficient" is considered as ``efficient". Although this problem can be solved, our focus here is to demonstrate a simple feasible ranking approach, we leave the natural language processing problem for future work. Other limitations to this proof of concept are posts that carry multiple code snippets, in which case all snippets get the same score. Further, some of these code snippets might not even implement the given topic. For instance, an SO post on calculating averages that is concerned about exceeding the datatype double's limits had a code snippet to explain double representation\footnote{http://stackoverflow.com/questions/1930454/what-is-a-good-solution-for-calculating-an-average-where-the-sum-of-all-values-e}.

While these limitations exist, our objective is to show that ranking based on desired properties of code snippet is feasible even with a simple approach. Figure~\ref{fig:RankingComparison} compares our results with the default output of CodeExchange.

\paragraph*{More applications} We believe this characterization of variants opens up several new applications. For instance, code completion can be sensitive to desired properties. Learning API usage can also benefit from this fundamental idea of looking at code snippets along with the desired properties. Often, when a better code is shown to the developers, they are able to understand the flaws in the existing code. Thus, this approach can help in bug detection as well.

%% file: 7_relatedwork.tex
\section{Related Work}
To the best of our knowledge, this is the first characterization study on code variants. However, there is work on independent aspects related to variants such as automatically mining relevant code~\cite{Keivanloo:2014:SWC:2568225.2568292,Terragni:2016:CAS:2931037.2931058,Sahavechaphan:2006:XMS:1167473.1167508}, and calculating similarity measures~\cite{Vinayakarao:2015:SHS:2678015.2682537,Mishne:2004:SCR:2816272.2816322} on source code snippets. There is also work on extracting code contexts~\cite{Nguyen:2012:GPC:2337223.2337232,Ponzanelli:2014:MST:2597073.2597077}. The studies on code clones~\cite{Lee:2010:ICC:1882291.1882317,Juergens:2009:CCM:1555001.1555062} are also close to our work. However, none of these works consider code variants as the first-class entity and investigate mining, clones, or context from that perspective.


\paragraph*{Searching/Mining Relevant Code Snippets} Researchers have largely ignored the search for ``better" source code snippets to replace an existing snippet. This requires ranking of snippets based on properties of interest to the developers. Existing research related to searching source code snippets focus on handling partial programs~\cite{Mishne:2012:TSC:2384616.2384689}, indexing~\cite{Lee:2010:ICC:1882291.1882317}, refining queries~\cite{Haiduc:2013:QQP:2486788.2486991}, and such aspects of building a search system. Majority of existing research which compares code snippets is limited to structural~\cite{Bajracharya:2010:LUS:1882291.1882316} and semantic similarity~\cite{DBLP:conf/dagstuhl/Koschke06}. 

\paragraph*{Extracting Code Context} Importance of understanding and extracting code context is highlighted in several existing works~\cite{Poulin:2014:TBM:2630768.2630776,Ponzanelli:2014:MST:2597073.2597077} as discussed in Section~\ref{ExecutionContext}. However, none of these papers study code context in detail. Our work builds upon their findings and provides a thorough review of different types of code contexts.
\paragraph*{Studies on redundancies in source code} Hindle et al.~\cite{Hindle:2012:NS:2337223.2337322} observe that source code being a human product is repetitive. Juergens et al.~\cite{Juergens:2010:CSB:1955601.1955971} claim that semantically similar code taken from various sources can be syntactically heterogeneous. In another study~\cite{Juergens:2009:CCM:1555001.1555062}, they also report that inconsistent changes to clones lead to maintenance issues. We leverage these studies, but find that semantic clones have inconsistent definition and code variants are conceptually different from them.

\paragraph*{Diversity in search results} Ambiguous queries are shown to demand diversity in search results~\cite{Welch:2011:SRD:1963405.1963441}. Martie and van der Hoek~\cite{Martie:2015:SEC:2820518.2820530} explore diversity in code search results. Our work differs from them because of the computation of diversity based on desired attributes. We believe that our work complements the existing work on computing diversity in code search results. Developers benefit from avoiding looking at semantically similar search results.

%% file: 8_threats.tex
\section{Threats to Validity}
\paragraph*{Internal Validity} Studying code variants on open source projects poses several challenges. We have used nine open source projects for evaluation. We strived to control variability by selecting 100 random bugs each from multiple projects. Other projects and domains might give different numbers. We have selected each project from a different domain to mitigate this threat. It is possible that these bugs are not representative of these projects. Also, these projects may not be representative of the entire set of open source projects. Choosing multiple projects was an attempt to reduce this kind of bias. However, we have ensured that there is no bias in terms of project size or popularity within our data set.  

\paragraph*{External Validity} Our results may not generalize to all types of code (for instance, scripting or functional). For evaluation, we have taken a mix of Java, C and Python projects popularly used in clone research. Hence, our work applies to high-level imperative languages at the least. Our results for variant mining depends on the discussion forum data. These results may not generalize to variants that are domain specific implementations with inadequate developer discussions.

%% file: 9_conclusion.tex
\section{Conclusion}
Code variants are very different from other structural and semantically similar code snippets such as code clones. They appear frequently in source code and developer discussions. Therefore, understanding code variants is important for software development and maintenance. Currently, there are inconsistent definitions of code snippets that are similar. To the best of our knowledge, we present the first study to characterize and distinguish code variants from other types of code that are similar to each other. In this work, we define code variants, classify them as simple and complex, and categorize them into three main types: algorithmic, diction and resource-oriented.

With the availability of code from open source projects and discussion forums, developers are increasingly turning to online sources for finding implementation that match specific desired behavior. Further, with ``big code" becoming accessible to the research community, an understanding of code variants and their characteristics can help build tools that provide automated code (variant) searches, including functionality such as, ranking code search results, mining source code variants, and recommender systems for variants. As a demonstration of the possible use of our characterization of code variants, we show how search results from a code search engine can be reordered based on a desired (implementation) property that a developer might have (e.g., speed of execution). This constitutes a first step towards an open research area where code variants can be mined from heterogenous sources (e.g., project code, issue tracker discussions, Q\&A forums) by using the code context and the desired behavior that a developer is looking for.

%% file: 9_references.tex
\bibliographystyle{unsrt}

\input{main.bbl}